\documentstyle[11pt,amssymbols,epsf]{article}
\newcommand{\be}{\begin{equation}}
\newcommand{\ee}{\end{equation}}
\newcommand{\ba}{\begin{array}}
\newcommand{\ea}{\end{array}}
\newcommand{\bea}{\begin{eqnarray}}
\newcommand{\eea}{\end{eqnarray}}

\renewcommand{\l}{\newline\null}

\newcommand{\lrar}{\longrightarrow}
\newcommand{\rar}{\rightarrow}
\newcommand{\p}{\partial}
\newcommand{\ol}{\overline}
\newcommand{\ti}{\tilde}
\newcommand{\la}{\langle}
\newcommand{\ra}{\rangle}
\def\figskip{\vskip .5cm plus 3mm minus 2mm}

\begin{document}
\begin{titlepage}
October, 1993\hfill PAR-LPTHE 93/50
\newline\null \hfill{\em hep-th yynnmmm}
\vskip 5cm\centerline
{\bf DYNAMICALLY BROKEN $\bf{U(1)_L}$ GAUGE THEORIES IN FOUR DIMENSIONS}
\vskip 1cm
\centerline{{B. Machet}
\footnote{Member of Centre National de la Recherche Scientifique}
\footnote{E-mail: machet@lpthe.jussieu.fr }}
\vskip 3mm
\centerline{{\em Laboratoire de Physique Th\'eorique et
Hautes \'Energies}
\footnote{LPTHE tour 16-1er \'etage, Universit\'e P. et M. Curie, BP 126,
4 place Jussieu, F 75252 PARIS CEDEX 05 (France).}, {\em Paris}}
\centerline{{\em Unit\'e associ\'ee au CNRS URA 280}}
 \vskip 2cm
{\bf Abstract:}  We study a dynamically broken $U(1)_L$ gauge theory endowed
with a composite scalar doublet (one scalar and one pseudoscalar); its
Lagrangian only differs from that of an abelian `Standard Model' by the
addition of a derivative coupling between a Wess-Zumino field, linked to the
previous scalars, and the fermionic current. Yet, in the Feynman path integral,
the non independence of the fermionic and scalar variables of integration
requires the introduction of constraints. When the gauge symmetry
is broken by the vacuum expectation value of the scalar field, they freeze all
degrees of freedom but those of a massive gauge field, including a (abelian)
pion. The anomaly disappears and the gauge current is conserved.
This is shown, and renormalizability studied, in the `Nambu-Jona-Lasinio
approximation'. Unitarity is demonstrated on general grounds.
\smallskip


\vfill
\null\hfil\epsffile{LOGO.eps}
\end{titlepage}


\section{Introduction.}

Our increasing understanding of the subatomic world, the success of the
Standard Model of electroweak interactions \cite{GlashowSalamWeinberg},
the compatibility of Quantum Chromodynamics  \cite{MarcianoPagels} with high
energy hadronic processes, cannot hide the
troubling persistence of fundamental problems. Whatever conceptual they may
appear, their relevance for practical matters is also important, and they
become more and more pressing as the cost of looking for hypothetical new
particles increases to hardly acceptable levels and will certainly not be
pursued without strong and unambiguous justifications. Among those
questions, those of importance for us here are, at the `experimental' level:\l
\quad$\ast$ the elusiveness of the Higgs boson;\l
\quad$\ast$ the non observation of the quarks as particles (they
apparently do not exist as asymptotic states but only as {\em fields} in the
Lagrangian), while observed {\em particles} (`pions' \ldots) do not appear in
the Lagrangian (this is actually not entirely the case when anomalies
\cite{AdlerBellJackiw} are involved, since a Wess-Zumino field
\cite{WessZumino}, which decays like a pion into two photons, has then to be
introduced for the Ward Identities to be satisfied by a local functional of
the fields.);\l
and, at a more conceptual level:\l
\quad$\ast$ the problem of anomalies in gauge theories, solved up to now by
a cancellation between quarks and leptons
\cite{BouchiatIliopoulosMeyer}, unsatisfying when one considers the
totally different nature of the two types of objects.

We have shown in \cite{BellonMachet1} how the leptonic sector of the
Standard Model can be made purely vectorial, and thus anomaly-free, without
any contradiction with experiment; the $V-A$ structure of weak currents and
the unobservability of the `right-handed' neutrino have been given the same
origin. We complete here the disconnection between the two sectors by
making a $U(1)_L$ hadronic model anomaly-free; its fermions are unobservable,
and so is the scalar boson responsible for the breaking of the symmetry.

Anomalies being central to our concern, we emphasize this point of view.
After the pioneering works \cite{BouchiatIliopoulosMeyer} showing that
achieving both unitarity and renormalizability was impossible in the strict
framework of a spontaneously broken gauge theory with a `standard' scalar
sector, `standard' meaning here a fundamental Higgs and its Goldstone
partner(s), the quest for gauge invariance was revived in the recent years.
It was shown  that the symmetry could be  restored with
the introduction of  extra scalar fields and their additional contributions
to the Lagrangian, involving in particular a Wess-Zumino term with a reversed
sign. First introduced by hands \cite{FaddevShatashvili}, it was then shown
\cite{BabelonSchaposnikViallet} that, in the functional integral formalism,
this appeared as a natural consequence of the integration over the whole
space of connexions and not only over the orbit space. However, though gauge
invariance was indeed recovered, it was soon recognized
\cite{BabelonMachetViallet} that the original problem of achieving both
unitarity and renormalizability was unsolved. The source of the difficulty
lies in that this degree of freedom (in the $U(1)$ case) appears in fact as a
dimensionless scalar field.  Power counting requires its
propagator to behave as $1/k^4$, which unavoidably reintroduces in the theory
an uncancelled ghost pole.  So, though no definitive proof has been established
yet, there is a widespread belief that the problem cannot be solved strictly
along these lines.

The step that we take here is to abandon the idea that the extra scalar field
$\xi$ (called hereafter the Wess-Zumino \cite{WessZumino} field) is an
independent degree of freedom; when the theory is spontaneously broken by
scalar fields $(H,\varphi)$ such that $\langle H\rangle = v \not= 0$, and
when the gauge generator and the unit matrix form an associative algebra,
there is a unique solution $\xi =\xi(H,\varphi)$ such that $\xi$ transforms
non-linearly by a gauge transformation on the fermion  fields if  $H$ and
$\varphi$ are themselves taken as composite fermion operators.
The field $\xi$ can be gauged into the third polarization of the massive gauge
boson, and we shall explicitly perform this transition to the `unitary gauge'.

The scalars and the fermions not being independent, integrating on both
types of variables in
the Feynman path integral requires introducing the appropriate constraints.
They can be transformed into an effective Lagrangian, which we treat in the
`Nambu-Jona-Lasinio approximation' \cite{NambuJonaLasinio}, akin to
propagating only bound sates in internal lines, and to truncating
perturbative series at the first power in an expansion in powers of $1/N$,
where $N$ is the number of flavours of fermions. We show that the latter
are given an infinite mass, making them unobservable, and yielding the
cancellation of the anomaly. The gauge current is conserved, achieving gauge
invariance. The scalar boson becomes unobservable too, and the effective
4-fermions couplings occurring in the Lagrangian of constraint turn out to
vanish, opening the way to renormalizability.

It is shown in \cite{Machet1} that the pseudoscalar composite
behaves like a (abelian) pion, having the usual couplings to leptons
and, despite the absence of anomaly, to two gauge fields \cite{Adler}.
No extra scale of interaction is {\em a priori} needed, unlike in
`technicolour' theories \cite{SusskindWeinberg}, as confirmed in the study of
the non-abelian case \cite{BellonMachet2}.

Using composite scalars unifies the two often disconnected phenomena of
gauge ($\la H\ra = v$) and chiral \cite{AdlerDashen}
($\la\ol\Psi \Psi\ra = N\mu^3$) symmetry breaking: the Goldstone(s) of the
broken symmetry is (are) the third polarization(s) of the massive gauge
field(s) {\em and} some precise linear combinations, controlled in the
non-abelian case by the Kobayashi-Maskawa \cite{KobayashiMaskawa}
mixing matrix, of the observed pseudoscalar mesons. This translates into
the expectation that those mesons which include a top quark should have
masses not very different from that of the $W$ or $Z$ gauge bosons (see
\cite{BellonMachet2}).

Because more phenomenological aspects have already been dealt with in
\cite{Machet1}, and will again  be in \cite{BellonMachet2} for the general
case,
 we rather put here the accent of the field theoretical aspect of this
$U(1)_L$ dynamically broken model.

\section{The Wess-Zumino field as a fermionic bound state.}

Let us consider a spontaneously broken $U(1)_L$ gauge theory. The explicit
Lagrangian will appear in the next section.  The gauge group $U(1)_L$ acts on
the fermions and  on the gauge field $\sigma_\mu$, which are coupled by
a $(V-A)$ law.
There are N fermions  with the same coupling (N is the number of `flavours')
described by an N-vector $\Psi$ on which acts the $U(N)_L\times U(N)_R$
chiral group. We embed $U(1)_L$ into $U(N)_L\times U(N)_R$ and,
consequently, take the generator of the `left' gauge group
\be {\Bbb T}_L={1-\gamma_5\over 2}\ {\Bbb T}\ee
as an $N\times N$ matrix.

$Tr \{{\Bbb T},{\Bbb T}\}{\Bbb T}$ can be non-vanishing and, so
\cite{GeorgiGlashow}, the fermionic
current
\be J_\mu^\psi = \overline\Psi\gamma_\mu {\Bbb T}_L\Psi =
{1\over 2}(V_\mu -  J _\mu ^5)\ee
can be anomalous.

We choose $\Bbb T$ to satisfy the condition
\be {\Bbb T}^2= 1,\label{eq:assoc}\ee
which is the simplest case when the gauge generator $\Bbb T$ and the unit
matrix
form an associative algebra.

We consider in the following the simple case where ${\Bbb T}$ is the unit
matrix
\be {\Bbb T}= 1_N,\ee
but other cases can be considered as well.

When eq.~(\ref{eq:assoc}) is satisfied, there exists a particular
two-dimensional representation of the gauge group:
\be \Phi=(H,\varphi)={v\over N\mu^3}(\overline\Psi  1 \Psi,
-i\,\overline\Psi\gamma_5 {\Bbb T}\Psi),\ee
where $H$ and $\varphi$ are scalar fields, both real.
Indeed, the action of $U(1)_L$ on $\Phi$ is deduced from that on $\Psi $ and
we have, using eq.~(\ref{eq:assoc}):
\be\begin{array}{c} {\Bbb T}_L\,.\,\varphi =i\,H,\\
     {\Bbb T}_L\,.\,H =-i\,\varphi.\end{array}\label{eq:gract}\ee
The necessity of relation (\ref{eq:assoc}) lies in that, because of the
$\gamma_5$ present in the action of the group, acting on composite scalars
will involve both commuting and anticommuting its generator(s); in this
simple $U(1)$ case, $\Bbb T$ of course commutes with itself while
eq.~(\ref{eq:assoc}) ensures that its anticommutator  with itself
is the unit matrix. This property will find its full justification in
the non-abelian case \cite{BellonMachet2} where we
show that, in the case of the Standard Model, the three $N\times N$ generators
${\Bbb T}^i$ of the $SU(2)$ group and the unit $N\times N$ matrix linked with
weak hypercharge form a matrix algebra. It is thanks to this peculiar algebraic
structure that we can build, there, a composite multiplet of four real scalar
fields.

The gauge theory is spontaneously broken by
\be \langle H\rangle=v,\ee
which is equivalent to
\be\langle\overline\Psi\Psi\rangle=N\mu^3.\ee
We call $H$ the scalar boson and $\varphi$ the pseudoscalar boson.
It is an immediate consequence of our construction that the so-called `gauge
breaking' and `chiral breaking' are the same phenomenon, the mechanism of
which we will not investigate here.

Writing
\be H= v+h,
\label{eq:shift1}\ee
let us define $\tilde H $ and $\xi$, both real too, by :
\be \tilde H =e^{-i\frac{\xi}{v}{\Bbb T}_L}.\ (H+i\varphi)\label{eq:chvar1}\ee
with
\be \tilde H = v+\eta.\label{eq:shift2}\ee
The solution of eq.~(\ref{eq:chvar1}) is
\be \left\{\ba{rcl}
0 &=& H\sin{\xi\over v}+\varphi\cos{\xi\over v}, \\
\tilde H &=& H\cos{\xi\over v}-\varphi\sin{\xi\over v},
\ea\right .\label{eq:chvar2}\ee
from which $\eta$ and $\xi$ can be expressed as series in $h/v$ and
$\varphi/v$:
\be\ba{lcl}
\xi &=& -\varphi\,(1-{h\over v} + {h^2\over v^2}- {\varphi^2 \over 3v^2} +
\cdots\ ),\\
\eta &=& h+{\varphi^2\over 2v}(1-{h\over v}) +\cdots .
\ea\label{eq:chvar3}\ee
The laws of transformation of $\tilde H$ and $\xi$ come from eqs.~
(\ref{eq:gract}) and (\ref{eq:chvar2}):
\be
when\qquad \Psi \lrar e^{-i\theta {\Bbb T}_L} \Psi,
\ee
\be\left\{\ba{lcl}
\xi &\lrar & \xi -\theta v,\\
\tilde{H} &=& invariant.
\ea\right .\label{eq:trans}\ee
A gauge transformation induces a translation on the field $\xi$,
equivalent to:
\be
e^{i{\xi\over v}{\Bbb T}_L} \lrar e^ {-i\theta {\Bbb T}_L}\
e^ {i{\xi\over v}{\Bbb T}_L}.
\ee
Eq.~(\ref{eq:trans}) corresponds to a non-linear realization of the gauge
symmetry \cite{WessWeinbergZumino}.
$\xi$ thus appears as a natural candidate for a Wess-Zumino field
\cite{WessZumino}.  In a gauge transformation, the variation of $\sigma _\mu$
gets cancelled by that of $-(1/g)\ \p_\mu \xi/ v$;  the generic gauge field
\be A_\mu = \sigma _\mu -{1\over g}{\p _\mu\xi\over v}
\label{eq:amu}\ee
will be now used to quantize the theory along the lines of
\cite{BabelonSchaposnikViallet}.

\section{Quantizing.}

Let the Lagrangian ${\cal L}(x)$ be at the start a function of $A_\mu$
 defined in eq.~(\ref{eq:amu}) above; this has been advocated in
\cite{BabelonSchaposnikViallet} to lead to the recovery of gauge invariance
for anomalous theories and thus to be the right choice in the process of
quantization:
\be\ba{ccl} {\cal L}(x)&=& -{1\over 4} F_{\mu \nu}F^{\mu\nu}
+i\overline\Psi \gamma^\mu\left(\p_\mu -ig(\sigma _\mu
-(1/ g)\ \p_\mu \xi/v)\,{\Bbb T}_L\right)\Psi\\ & &
+{ 1\over 2} \p_\mu \tilde H \p^\mu \tilde H
+{1\over 2} g^2\left(\sigma _\mu -(1/g)\ \p_\mu \xi/v\right)^2
\tilde H^2 - V(\tilde H^2). \ea\label{eq:L1}\ee
$V(\tilde H^2)$ is the scalar potential, a polynomial of degree 4 in $\tilde
H$.
We shall see that it is finally `screened' by the constraints.
$\cal L$ has the (classical) invariance
\be
\left\{\ba{lcl}
\xi &\lrar & \xi -\theta v,\\
\sigma_\mu &\lrar & \sigma_\mu -(1/g)\  \p_\mu \theta.
\ea\right .
\label{eq:clasym}\ee
{}From the definition of $\tilde H$ and $\xi$ in eq.~(\ref{eq:chvar1}), we have
\be
\ti H^2 = H^2 +\varphi^2,
\ee
and
\be
{1\over 2} \p_\mu \tilde H \p^\mu \tilde H +{1\over 2}g^2(\sigma_\mu
-{1\over g}\p_\mu{\xi\over v})^2\tilde H^2 =
{1\over 2}(D_\mu H D^\mu H +D_\mu\varphi D^\mu\varphi),
\label{eq:Ls1}\ee
with
\be\ba{lclcl}
D_\mu H &=& \p_\mu H -ig\sigma_\mu {\Bbb T}_L.H
                       &=& \p_\mu H -g\sigma_\mu\varphi, \\
D_\mu\varphi &=& \p_\mu\varphi -ig\sigma_\mu {\Bbb T}_L.\varphi
                       &=& \p_\mu\varphi +g\sigma_\mu H,
\ea\ee
from which eq.~(\ref{eq:L1}) transforms into a more customary form for $\cal
L$,
in terms of $H$ and $\varphi$:
\be\ba{lcl}
{\cal L} &=&
-{1\over 4}F_{\mu\nu}F^{\mu\nu} +i\overline\Psi
\gamma^\mu(\p_\mu -ig\sigma_\mu {\Bbb T}_L)\,\Psi\\
& &
+{1\over 2}\left((D_\mu H)^2 +(D_\mu \varphi)^2\right) -V(H^2 +\varphi^2)\\
& & -{1\over v}\p^\mu\xi(h,\varphi)\  J_\mu^\psi,
\ea\label{eq:L2}\ee
where we recognize an abelian `Standard Model' to which has been added the
({\em a priori} non-renormalizable) derivative coupling (after integrating by
parts)
\be
(\xi/ v)\ \p^\mu J_\mu^\psi,
\label{eq:WZ}\ee
of the Wess-Zumino field $\xi$ (expressed as a function of $h$ and $\varphi$)
 to the fermionic current. We recall that its presence is the consequence of
choosing a Lagrangian function of $A_\mu$.

Would not the scalar fields be composite, eq.~(\ref{eq:L1}) would describe a
theory of a massive $A_\mu$, unitary, but with a  non-renormalizable coupling
$\tilde H^2 (\p_\mu\xi/v)^2$. However, $H$ and $\varphi$ being `made up'
with fermions, the Feynman integration on both types of variables can only be
performed at the price of introducing constraints
\be
\prod_x \delta(C_H(x))\prod _x \delta (C_\varphi(x))
\ee
with
\be
C_H= H-{v\over N\mu^3}\ol\Psi\Psi,
\label{eq:CH}\ee
\be
C_\varphi=\varphi +i\,{v\over N\mu^3}\ol\Psi\gamma_5{\Bbb T}\Psi.
\label{eq:Cphi}\ee
We thus define the theory by
\be Z= \int {\cal D} \Psi{ \cal D} \overline\Psi{\cal D}H
{\cal D} \varphi {\cal D} \sigma_\mu \  e^{i\int d^4 x {\cal L}(x)}
\prod _x \delta(C_H(x))\prod _x \delta (C_\varphi(x)).\label{eq:Z} \ee
Note that the constraints are no longer invariant by the transformation
(\ref{eq:clasym}), but by acting on both fermions and scalars like in eq.
(\ref{eq:chvar}) below. Rewriting the
$\delta$ functionals in their exponential form, we transform them into the
effective Lagrangian, that we shall call Lagrangian of constraint
\be
{\cal L}_c =
\lim_{\beta\rar 0}\frac{-N\Lambda^2}{2\beta}
\left(H^2 +\varphi^2 -{2v\over N\mu^3}(H\overline\Psi\Psi-i\,\varphi
\overline\Psi\gamma_5 {\Bbb T} \Psi)+
 {v^2\over N^2\mu^6}\left((\overline\Psi\Psi)^2
-(\overline\Psi\gamma_5 {\Bbb T} \Psi)^2\right)\right).\label{eq:Lc1}
\ee
$\Lambda$ is an arbitrary mass scale.
Using eq.~(\ref{eq:chvar2}), ${\cal L}_c$ can also be written in terms of
$\tilde H$ and $\xi$
\be
{\cal L}_c =
\lim_{\beta\rar 0}\frac{-N\Lambda^2}{2\beta}
\left(\tilde H^2-{2v\over N\mu^3}\tilde H(\overline\Psi\Psi \cos{\xi\over v}
+i\overline\Psi\gamma_5 {\Bbb T} \Psi \sin{\xi\over v})+
{v^2\over N^2\mu^6}\left((\overline\Psi\Psi)^2
-(\overline\Psi\gamma_5 {\Bbb T} \Psi)^2\right)\right).\label{eq:Lc2}
\ee
{\em Remark 1}: to ease the computations we have exponentiated the two
constraints
on $H$ and $\varphi$ with the same coefficient $\beta$.

{\em Remark 2}: the integration over $\xi$ may be conceptually preferred to
that
over $\varphi$ as equivalent to integrating over the gauge group
\cite{BabelonSchaposnikViallet}. As we shall see further, they only differ
by a $\xi$ independent Jacobian.

\subsection{Unitarity.}

In a first step, we shall show how the introduction of the coupling
(\ref{eq:WZ}) cures, at tree level, the unitarity problem for anomalous
gauge theories, as exposed in the works \cite{BouchiatIliopoulosMeyer}.
This will be done forgetting the constraints. Then we shall give a general
argument close to that of Abers and Lee \cite{AbersLee} showing how, now in the
constrained theory, the Wess-Zumino field can be gauged into the third
component of the massive gauge boson and the transition to the `unitary gauge'
be performed.

\subsubsection{Tree level unitarity.}

Leaving aside the constraints, there is a first, obvious, way to show that the
theory is unitary, which is by going to the variable $A_\mu$ in
eq.~(\ref{eq:L1}).  The gauge field
becomes massive when $\tilde H$ gets a non-vanishing vacuum expectation
value and no ghost-like pole appears in any propagator.

But it is also instructive to directly answer the argument of ref.
\cite{BouchiatIliopoulosMeyer} in the Landau gauge, at tree level. We stay now
with the $\sigma_\mu$ variable. Together with the mass of the gauge field, a
kinetic term for $\xi$ is produced. $\xi$ is a Goldstone particle. The
Landau gauge is specially convenient as the non-diagonal coupling
$\sigma_\mu \p^\mu\varphi$ plays no more role. The gauge field propagator is
then ($M=gv$)
\be
D_{\mu\nu}^\sigma = -i\left(\frac{g_{\mu\nu}-k_\mu k_\nu/k^2}{k^2 -M^2}\right),
\ee
which has the usual ghost pole at $k^2=0$. Since $\sigma_\mu$ is coupled to a
potentially non-conserved current, this can break unitarity, unless the
corresponding residue is cancelled by that of another massless particle.
This is precisely the role that $\xi$ plays here, as shown on fig.~1.
\figskip
\hskip -0.5truecm
\epsffile{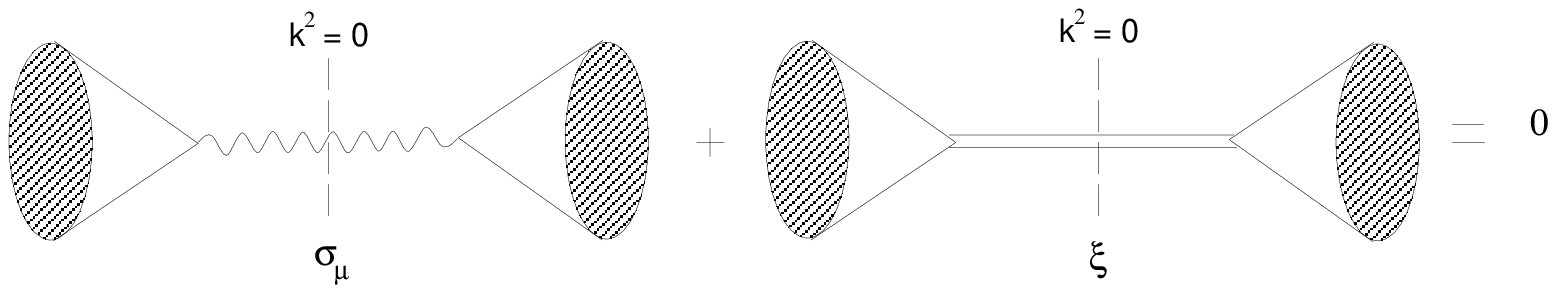}
\figskip
\centerline {\em Fig.1: Cancellation of the ghost residue by the Goldstone
in the Landau gauge.}
\figskip
In other $\alpha$-gauges corresponding to a gauge-fixing Lagrangian ${\cal
L}_{gf} = -1/2\alpha\,(\p_\mu\sigma^\mu)^2$, there exists a non-diagonal
coupling
\be
-M \sigma_\mu\p^\mu\xi.
\ee
The resummed gauge field propagator (see fig.~2) is
\be
\tilde D_{\mu\nu}^\sigma = -i\left(\frac{g_{\mu\nu}-k_\mu
k_\nu/k^2}{k^2 -M^2}+\alpha \frac{k_\mu k_\nu}{k^4}\right),
\ee
and the resummed $\xi$ propagator (see fig.~2)
\be
\tilde D^\xi = i\ \frac{k^2 -\alpha M^2}{k^4}.
\ee
%
%
\figskip
\hskip -1truecm
\epsffile{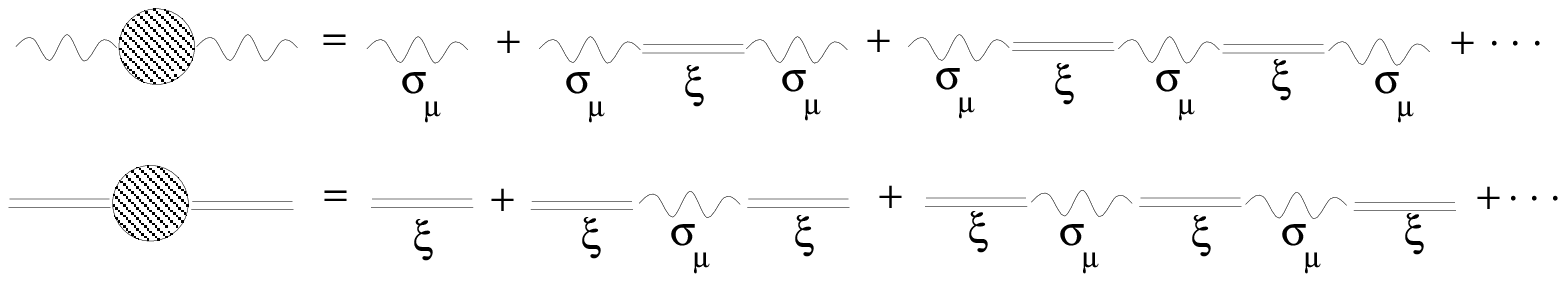}
\figskip
\centerline {\em Fig.2: Resumming the $\sigma_\mu$ and $\xi$ propagators.}
\figskip

In addition to the diagrams of fig.~1, we have now to look also at those of
fig.~3  and we check that the cancellation of the residue of the ghost pole
again occurs.
\figskip
\epsffile{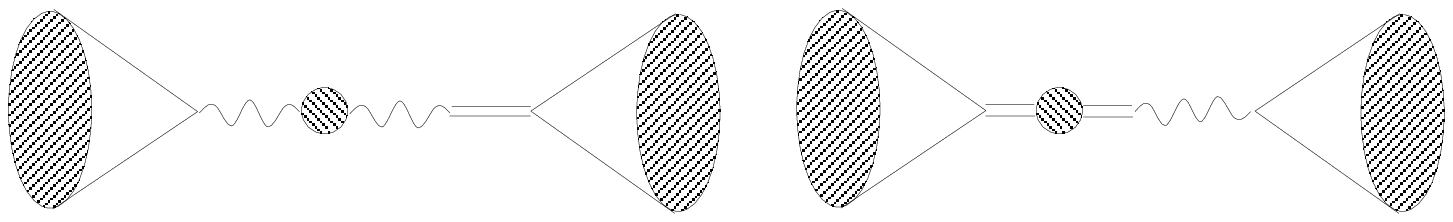}
\figskip
\centerline {\em Fig.3: Other diagrams for ghost cancellation in a general
$\alpha$-gauge.}
\figskip
This explicitly shows the role of $\xi$ in restoring unitarity through its
derivative coupling to the fermionic current.

\subsubsection{The general argument.}

Considering the full (constrained) theory, we first perform in $Z$ the
change of variables ($\chi$ being a function of $x$)
\be
\left\{\ba{rcl}
\Psi &\lrar & e^{-i(\chi/v){ \Bbb T}_L}\Psi,\\
H+i\varphi &\lrar & e^{-i(\chi/v) {\Bbb T}_L}.\ (H+i\varphi);
\ea\right .
\label{eq:chvar}\ee
it leaves ${\cal L}_c$ invariant and yields two Jacobians:\\
-  the first, coming from the transformation of the fermionic measure
\cite{Fujikawa79}, is
\be
J = e^{i\int d^4x {(\chi/v){\cal A}}},
\ee
where $\cal A$ is the (eventual) anomaly;\\
-  the second, corresponding to a `rotation' of the scalars, is unity.

 We use the laws of transformation
(\ref{eq:trans}) and the fact that the scalar Lagrangian ${\cal L}_s$
\be\ba{lcl}
{\cal L}_s &=& { 1\over 2} \p_\mu \tilde H \p^\mu \tilde H
+{1\over 2} g^2\left(\sigma _\mu -(1/g)\ \p_\mu \xi/v\right)^2
\tilde H^2 - V(\tilde H^2)\\
&=&
{1\over 2}(D_\mu H D^\mu H +D_\mu\varphi D^\mu\varphi) -V(H^2 + \varphi^2)
\ea\label{eq:Ls2}\ee
is invariant when one transforms both the gauge field and the scalars:
\be
{\cal L}_s(\xi-\theta v, \tilde H,\sigma_\mu) =
{\cal L}_s\left(\xi, \tilde H, \sigma_\mu +(1/g)\,\p_\mu \theta\right)
\ee
to deduce that, by the change of variables (\ref{eq:chvar}), one gets an
effective Lagrangian
\be\ba{lcl}
{\cal L}'+ {\cal L}_c &=&
-{1\over 4}F_{\mu\nu}F^{\mu\nu} +i\overline\Psi
\gamma^\mu(\p_\mu -ig\sigma_\mu {\Bbb T}_L)\Psi\\
& &
+{1\over 2}\left(
\left(\p_\mu H -g(\sigma_\mu +{1\over g}{\p_\mu \chi\over v})\varphi\right)^2
+\left(\p_\mu \varphi +g(\sigma_\mu +{1\over g}{\p_\mu \chi\over v})H\right)^2
\right)
-V(H^2 +\varphi^2)\\
& & -(\chi/ v)\  (\p^\mu J_\mu^\psi -{\cal A}) +{1\over v}
(\xi-\chi)\,\p^\mu J_\mu^\psi \\
& & + {\cal L}_c.
\ea\label{eq:Lprime}\ee
Some explanations are in order:\l
\quad - the first $-(\chi/v)\,\p^\mu J_\mu^\psi$ comes from the transformation
of the fermions;\l
\quad - the $(\chi/v)\,{\cal A}$ comes from $J$;\l
\quad - the $1/v(\xi - \chi)\,\p^\mu J_\mu^\psi$ is the original derivative
coupling (\ref{eq:WZ}) after $\xi$ has been shifted by $\chi$ by the action of
(\ref{eq:chvar}) according to (\ref{eq:trans}).

We then choose $\chi = \xi$ and finally go to the integration variables
$\xi$ and $\tilde H$ defined by eqs.~(\ref{eq:chvar1}) and (\ref{eq:chvar2}).
This yields one more Jacobian $J_1$ according to
\be
{\cal D} H {\cal D} \phi = J_1\  {\cal D} \tilde H {\cal D}
\xi,
\ee
which can be expressed as
\be
J_1 =\prod _x {\ti H(x)\over v}=
\exp \left(\delta^4(0)\int d^4x\sum_{n=1}^{\infty}
{(-1)^{(n+1)}}{({\eta/v)^n}\over n}\right).
\label{eq:J1}\ee
Finally, after the two transformations above, Z becomes
\be
Z = \int {\cal D} \Psi{ \cal D} \overline\Psi
{\cal D} \tilde H {\cal D} \xi {\cal D} \sigma_\mu \  J_1 \
e^{i\int d^4 x \left(\ti{\cal L}(x)+ {\cal L}_c(x)\right)},
\label{eq:Ztilde}\ee
with, using again eq.~(\ref{eq:Ls1}),
\be\ba{ccl}
\ti{\cal L}& =&
-{1\over 4}F_{\mu\nu}F^{\mu\nu} +i\overline\Psi\gamma^\mu
\left( \p_\mu -ig\sigma_\mu {\Bbb T}_L\right)\Psi
+(\xi/v)\,{\cal A}-(\xi/v)\,\p^\mu J_\mu^\psi\\
& &
+{1\over 2}\p_\mu\tilde H\p^\mu\tilde H +{1\over 2}\sigma_\mu^2 \tilde H^2
-V(\tilde H^2).
\ea\label{eq:Ltilde}\ee
As $J_1$ in eq.~(\ref{eq:J1}) does not depend on $\xi$, the $\xi$ equation
coming from eqs.~(\ref{eq:Ztilde}) and (\ref{eq:Ltilde}) is now
\be
\p^\mu J_\mu^\psi -{\cal A} - v{\p{\cal L}_c\over \p\xi}=0.
\label{eq:xieq}\ee
$v\ {\p{\cal L}_c /\p\xi}$ is the classical contribution
$\p_\mu J^\mu_{\psi c}$, coming from ${\cal L}_c$, to the divergence of the
fermionic current, as can be seen by varying the Lagrangian
${\cal L} + {\cal L}_c$ with a global fermionic transformation, such that
eq.~(\ref{eq:xieq}) rewrites
\be
\p^\mu J_\mu^\psi -{\cal A} -\p^\mu J_{\mu c}^\psi =0,
\ee
and is nothing more than the exact  equation for $\p^\mu J_\mu^\psi$.
We have thus put $Z$ in a form where $\xi$ appears as a Lagrange multiplier;
the associated constraint being satisfied at all orders, $\xi$ disappears:
it has been `gauged away' and transformed into the third polarization of the
massive vector boson.  This form of the theory describing a massive gauge
field is manifestly unitary. The transformations performed are equivalent to
going to the `unitary gauge'.

{\em Remark}: it is likely that all three polarizations of the massive gauge
field
are composite. This point, likely to be linked with the physics of vector
mesons, will not be investigated here.

\section{The Nambu-Jona-Lasinio approximation.}

The theory defined by
\be
Z =  \int {\cal D} \Psi{ \cal D} \overline\Psi{\cal D}H
  {\cal D} \phi {\cal D} \sigma_\mu \  e^{i\int d^4 x ({\cal L} + {\cal
L}_c)(x)}
\ee
can be thought {\em a priori} pathological because of the presence of
non-renormalizable couplings:\l
\quad - the 4-fermions couplings of ${\cal L}_c$;\l
\quad - the derivative coupling of the Wess-Zumino field to the fermionic
current, eq.~(\ref{eq:WZ}).\l
We shall however see that in the approximation of resumming ladder diagrams
of 1-loop fermionic bubbles or, equivalently, of dropping contributions at
order higher than $1$ in an expansion in powers of $1/N$ (Nambu-Jona-Lasinio
approximation \cite{NambuJonaLasinio}), special properties are
exhibited:\l
\quad - the effective 4-fermions couplings go to $0$ and the effective
fermion mass goes to infinity;\l
\quad - the scalar $h$ (or $\eta$) also decouples;\l
\quad - the gauge anomaly disappears and the gauge current is conserved.

In this approximation, our model will be shown to be a gauge invariant,
anomaly-free theory, the only asymptotic states of which are the three
polarizations of the massive gauge field (one of them being the composite
field $\xi$, shown in \cite{Machet1} to behave like an abelian pion).
Isolated fermions are no longer observed as `particles', showing the
consistency of this approximation, known to propagate only fermionic bound
states \cite{NambuJonaLasinio}.

The analysis being based on truncating an expansion in powers of $1/N$,
we make precise our counting rules:\l
\quad - $g^2$ is of order $1/N$ (see \cite{Coquereaux});\l
\quad - the 4-fermions couplings are of order $1/N$ (see eq.~(\ref{eq:Lc1}));\l
\quad - from the definitions of $h$ and $\varphi$ ($\eta$ and $\xi$), their
propagators are also of order $1/N$, a factor $N$ coming from the associated
fermionic loop;\l
\quad - thus, we shall consistently attribute a power $N^{-1/2}$ to the
fields $h, \varphi, \eta, \xi$, and $N^{1/2}$ to fermions bilinears
including a sum over the flavour index, such that, as expected,
$g\,\sigma^\mu\,\overline\Psi\gamma_\mu{\Bbb T}_L\Psi$ is of order $1$ like
$\sigma_\mu$ itself and the whole Lagrangian $\cal L$.

\subsection{The effective fermion mass and 4-fermions couplings.}

For the sake of completeness, we reproduce the argumentation presented in
\cite{Machet1}.
At the classical level, the infinite fermion mass $m_0 = -\Lambda^2 v^2
/\beta\mu^3$ in ${\cal L}_c$
is cancelled by the 4-fermions term $\propto(\ol\Psi\Psi)^2$ when
$\la\ol\Psi\Psi\ra = N\mu^3$; however, staying in the above
approximation, the effective 4-fermions coupling $\zeta(q^2)$ and the
fermion mass $m$ satisfy the two coupled equations
\bea
\zeta(q^2) &=& \frac{\zeta_0}{1-\zeta_0 \; A(q^2, m)},\cr
\noalign{\vskip 2mm}
m &=& m_0 -2\zeta(0)N\mu^3,
\eea
graphically depicted in fig.~4 and fig.~5. $\zeta _0$ ($\zeta_0^5$) are the
bare 4-fermions couplings $\zeta _0 = - \zeta _0^5 = m_0/2N\mu^3$; $A(q^2,m)$
is the one-loop fermionic bubble. The above cancellation represents only the
first two terms of the series depicted in fig.~5.
\figskip
\pagebreak
\epsffile{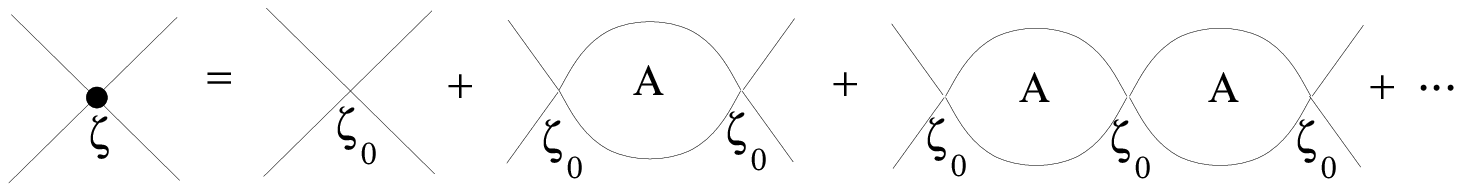}
\figskip
\centerline {\em Fig.4: The effective 4-fermion coupling $\zeta(q^2)$.}
\figskip
\hskip -.5cm\epsffile{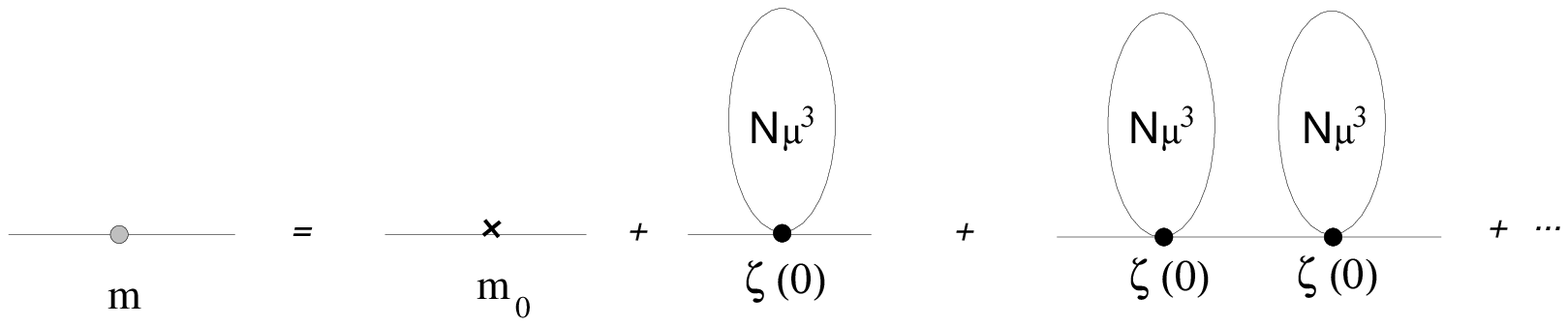}
\figskip
\centerline {\em Fig.5: Resumming the fermion propagator.}
\figskip
$\mu^3$ being finite, $m=m_0$ is a solution of the equations above
as soon as $\zeta(0)$ goes to $0$.
This is the case here since $\zeta(0)\propto -A(0,m)^{-1}$, and $A$ involves
a term proportional to $m^2$.
This also makes the effective 4-fermions coupling $\zeta(q^2)$, behaving
like $- A(q^2, m)^{-1}$, (and similarly $\zeta^5(q^2)$) go to $0$ like
$\beta^2$.

\subsection{The scalar field.}

The scalar potential, usually chosen as
\be
V(H,\varphi)= -{\sigma^2\over 2}(H^2+\varphi^2)+
{\lambda\over 4}(H^2+\varphi^2)^2
\ee
is modified by the constraint in its exponentiated form, to become
\be\ba{lcl}
\hskip -0.5cm\ti V(H,\varphi)&=& V(H,\varphi)\\
&+&\lim_{\beta\rar 0}
{N\Lambda^2\over 2\beta}\left(H^2 +\varphi^2 -
{2v\over N\mu^3}(H\overline\Psi\Psi-i\,\varphi \overline\Psi\gamma_5
{\Bbb T} \Psi)
+ {v^2\over N^2\mu^6}\left((\overline\Psi\Psi)^2
-(\overline\Psi\gamma_5 {\Bbb T} \Psi)^2\right)\right).
\ea\label{eq:Vtilde}\ee
Its minimum still corresponds to $\langle H\rangle = v$,
$\langle \overline\Psi \Psi \rangle =N\mu^3$, $\langle \varphi\rangle
= 0$ if $\sigma^2 =\lambda v^2$, but the scalar mass squared has now become
\be
\left .\frac{\p^2\ti V}{\p H^2}\right|_{H=v}= -\sigma^2 + 3\lambda
v^2 +{N\Lambda^2\over\beta},
\ee
which goes to $\infty$ at the limit $\beta\rar 0$ when the constraints are
implemented.

The coupling between the scalar and the fermions present in ${\cal L}_c$
does not modify this result qualitatively. Indeed, resumming the series
depicted in fig.~6, the scalar propagator becomes
\be
D_h = {\displaystyle\frac{D^0_h} {1-D^0_h
  \left(\frac{i\Lambda^2 v}{\beta\mu^3}\right)^2 A(q^2,m)}}\label{eq:Dh},
\ee
where $D^0_h$ is the bare scalar propagator
\be
D^0_h = \frac{i}{q^2 -{N\Lambda^2\over\beta}}.
\ee
At high $q^2$, $A(q^2,m)$ behaves like $N\,(a\,q^2 + b\,  m^2 + \ldots\,)$
(see for example \cite{Broadhurst}), such
that, when $\beta \rar 0$, $D_h$ now gets a pole at $q^2 = -(b/a)\,m^2$.
Checking \cite{Broadhurst} that the sign of $-b/a$ is positive confirms
the infinite value of the mass of the scalar. Furthermore, in this same
limit, $D_h$ goes to $0$ like $\beta^2$.

We conclude that the scalar field $h$ (nor, similarly, $\eta$)
cannot be produced either as an asymptotic state.
\figskip
\epsffile{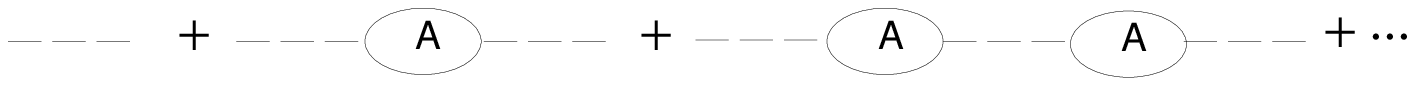}
\figskip
\centerline {\em Fig.6: Resumming the scalar propagator.}
\figskip
{\em Remark:} the constraints give $\xi$ (and $\varphi$) the same infinite mass
as $h$. The $\varphi$ resummed propagator $D_\varphi$ also vanishes like
$\beta^2$. However, unlike $\xi$, $h$ cannot be reabsorbed into the massive
gauge boson.

\subsection{Counting the degrees of freedom.}

It is instructive, at that stage, to see by which mechanism the degrees
of freedom have been so drastically reduced, since neither the scalar field
nor the fermions are expected to be produced as asymptotic states.
We started with 2 (2 scalars) + 4 (one vector field) + 4N (N fermions) degrees
of freedom. They have been reduced to only 3, the three polarizations of the
massive vector boson by 4N + 3 constraints which are the following:\\
- the 2 constraints linking $\varphi$ and $H$ to the fermions;\\
- the gauge fixing needed by gauge invariance;\\
- the 4N constraints coming from the condition
$\langle\overline\Psi\Psi\rangle =N\mu^3$: indeed because of the
underlying fermionic $O(N)$ invariance of the theory, this condition
is equivalent to
\be
 \langle\overline\Psi_n\Psi_n\rangle =\mu^3,
\  for\  n=1 \ldots\  N,
\label{eq:psi1}\ee
itself meaning (see \cite{NovikovShifmanVainshteinZakharov})
\be
 \langle\overline\Psi ^\alpha_n\Psi^\alpha_n\rangle =\mu^3/4,
\  for\  n=1 \ldots\  N\  and \ \alpha\ =1\ \ldots\  4,
\label{eq:psi2}\ee
which make 4N equations.

{\em Remark:} eq.~(\ref{eq:psi1}) does not imply eq.~(\ref{eq:psi2}); only the
reverse is true, and we take, according to
ref.~\cite{NovikovShifmanVainshteinZakharov}, the latter as a definition of
the former.

One should not conclude that the `infinitely massive' fermions play no
physical role \cite{ApplequistCarrazone}. Indeed, as shown below, they make
the anomaly disappear and, as studied in \cite{Machet1}, also trigger the usual
decays of the pion into two gauge fields.

\subsection{The disappearance of the anomaly and the conservation of the
gauge current.}

This theory with `infinitely massive' fermions has no anomaly.
It can most easily
be seen in the Pauli-Villars regularization of the triangular diagram, which
yields the covariant form of the anomaly. $M_{Reg}$ being the mass of the
regulator, the Ward Identity corresponding to fig.~7 writes
\be
k^\mu \Big(T_{\mu\nu\rho}(m) - T_{\mu\nu\rho}(M_{Reg})\Big) =
m T_{\nu\rho}(m) -M_{Reg} T_{\nu\rho}(M_{Reg}).
\label{eq:Ward}\ee
We have
\be
\lim _{M_{Reg}\rar\infty} M_{Reg} T_{\nu\rho}(M_{Reg}) =
-{\cal A}(g,\sigma_\mu),
\ee
where ${\cal A}(g,\sigma_\mu)$ is the anomaly; so,  when $m\rar\infty$,
the Ward Identity (\ref{eq:Ward}) now shows that the anomaly gets cancelled.
\figskip
\hskip 2cm\epsffile{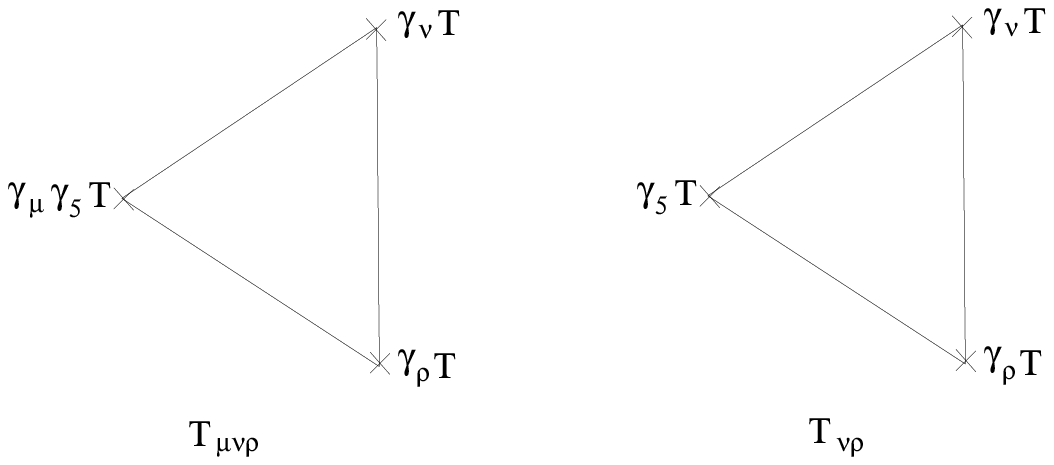}
\figskip
\centerline {\em Fig.7: Triangular diagrams involved in the anomalous Ward
Identity.}
\figskip
This is exactly the inverse of the situation described in \cite{D'HokerFarhi}:
here, by decoupling, the fermions generate an effective Wess-Zumino term
exactly cancelling the anomaly initially present.

The importance of the large fermion mass limit and its relevance to the
low energy or soft momentum limit has also been emphasized in
\cite{Fujikawa86} in the case of the non-linear $\sigma$-model. This case is
all the more relevant as our scalar boson has been shown
to get itself an infinite mass.

Though the Lagrangian of the theory with constraints is no longer invariant by
(\ref{eq:clasym}), the gauge current $J_\mu^\sigma$ is conserved. It writes
\be\ba{lcl}
J_\mu^\sigma &=& g\,J_\mu^\psi + g^2\,\sigma_\mu(H^2+\varphi^2)
-g\,(\varphi\p_\mu H -H\p_\mu \varphi)\\
            &=& g\,J_\mu^\psi + g^2\, \tilde H^2 (\sigma_\mu - {1\over g}
{\p_\mu\xi\over v}).
\ea \label{eq:Jsigma}\ee
Using the invariance of ${\cal L}_c$ by a transformation acting on both
scalars and fermions (\ref{eq:chvar}) to transform the l.h.s. into a
variation with respect to $\xi$, the $\Psi$ equation yields
\be
\p^\mu J_\mu^\psi = v {\p{\cal L}_c\over \p\xi},
\label{eq:dJpsi}\ee
while the $\xi$ equation, deduced from the Lagrangian $\cal L$
(eq.~(\ref{eq:L1})) $+ {\cal L}_c$ (eq.~(\ref{eq:Lc2})), gives
\be
\p^\mu J_\mu^\sigma = -gv\, {\p{\cal L}_c\over \p\xi},
\label{eq:dJsigma}\ee
such that we have, from eqs.~(\ref{eq:dJsigma}) and (\ref{eq:dJpsi}),
\be
\p^\mu J_\mu^\sigma = -g\, \p^\mu J_\mu^\psi.\label{eq:dJ}
\ee
Now we can also make use of the $\xi$ equation obtained from the Lagrangian
 (\ref{eq:L1}) written as a function of $A_\mu$ which writes simply,
as $\xi$ now only appears in ${\cal L}_c$
\be
 {\p{\cal L}_c\over \p\xi} = 0.
\ee
 This entails, by eq.~(\ref{eq:dJpsi}), the conservation of
the fermionic current and, by eq.~(\ref{eq:dJ}), that of the gauge current
$J_\mu^\sigma$. In the absence of anomaly, this classical equation stays
valid at the quantum level, making exact the conservation  of the gauge
current. This implements gauge invariance in the constrained theory.

\subsection{Renormalizability.}

The two obstacles to renormalizability are the derivative coupling
(\ref{eq:WZ}) and the 4-fermions couplings of ${\cal L}_c$. We have shown
that we can go from the latter to effective couplings $\zeta(q^2)$
vanishing like $\beta ^2$. The reshuffled perturbative expansion built with
those effective couplings has the right properties for renormalizability
since:\l
\quad - all possible counterterms that could be expected in the
renormalization of the 1-loop diagrams depicted in fig.~8 and which are not
initially present in the Lagrangian, corresponding to 6-fermions,
8-fermions, 4 or 6-fermions-gauge field or 4- or 6-fermions-scalar interactions
can be dropped because of their higher order in $1/N$; furthermore, they all
go to $0$ like powers of $\beta$;\l
\figskip
\hskip2cm\epsffile{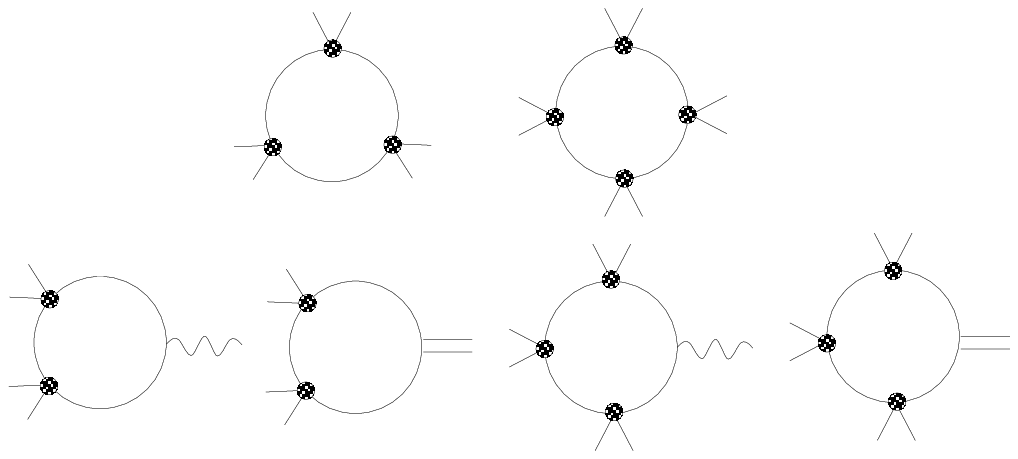}
\figskip
\centerline {\em Fig.8: Possible contributions to counterterms that can be
dropped.}
%
\quad - the diagram of fig.~9, potentially responsible of double counting,
also vanishes like \hbox {powers of $\beta$.}\l
\figskip
\hskip 4.5cm\epsffile{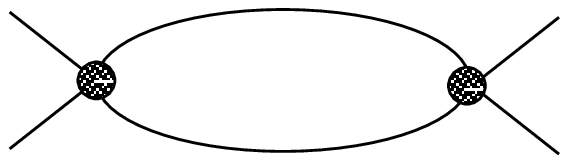}
\figskip
\centerline {\em Fig.9: Diagram that could cause double-counting problems.}
\figskip
The first obstacle has thus disappeared.

The second can also be eliminated thanks to the presence of the non-diagonal
$-M \sigma^\mu\,\p_\mu\xi$ coupling in the Lagrangian for the scalars. The
massive gauge field propagator being
\be
D_{\mu\nu}^\sigma = -i\,\frac{g_{\mu\nu} -k_\mu k_\nu/M^2}{k^2 - M^2},
\ee
the diagram depicted in fig.~10 exactly cancels the $(\xi/v)\,\p^\mu
J_\mu^\psi$ coupling.
\figskip
\hskip 4.5cm\epsffile{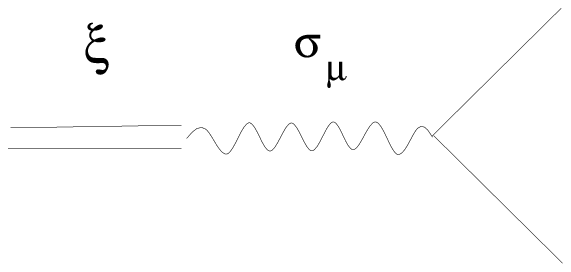}
\figskip
\centerline {\em Fig.10: Cancelling the derivative coupling of $\xi$ to the
fermions.}
\figskip
The problem can then be thought, in the $h,\varphi$ variables, to be
transferred to that of a $M\xi(h,\varphi)\,\p^\mu\sigma_\mu$ coupling, non
renormalizable either because $\xi$ is an infinite series in $h$ and
$\varphi$. However:\l
$\ast$
\be
M\xi(h,\varphi)\,\p^\mu\sigma_\mu =
          gv\,\varphi\,\p^\mu\sigma_\mu(1+{\cal O}(1/N));
\ee
$g$ being itself ${\cal O}(N^{-1/2})$, the correction can be neglected;\l
$\ast$ the (resummed) propagators $D_h$ and $D_\varphi$ vanishing like
$\beta^2$ (see section 4.2), the additional internal lines of $h$ and
$\varphi$ resulting from
taking $\xi(h,\varphi)$ instead of $\varphi$ in the above coupling would make
the corresponding diagrams vanish like powers of $\beta$.

We consequently end up with a renormalizable coupling.

{\em Remark}: the above is to be compared with the Standard Model, in which the
gauge fixing condition is chosen so as to precisely cancel the non-diagonal
coupling between the gauge fields and the Goldstones \cite{AbersLee}.

Another, maybe simpler, point of view, is to rewrite the full Lagrangian as
a function of $A_\mu$, thus reabsorbing the Wess-Zumino field into the gauge
field. Now, all couplings except the kinetic term for $\eta(h,\varphi)$
are renormalizable (the 4-fermions couplings being already dealt with).
However, for the same reasons as above, we can replace the kinetic term for
$\eta$ by one for $h$, and also replace $(1/2)g^2(H^2 + \varphi^2)A_\mu^2$
by $(1/2) g^2v^2\,A_\mu^2$. We thus describe a theory of a
massive $A_\mu$ coupled to the exactly conserved fermionic current
$J_\mu^\psi$. It can be `gauge fixed' (though this expression is abusive
here) and is renormalizable \cite{Stueckelberg}. This form of the theory is
extremely simple
since $h$ decouples, the equation for $\varphi$, no-longer a dynamical
field, just corresponds to its definition in terms of fermions; the latter
only play a role to cancel the anomaly (see also \cite{Machet1} and
\cite{BellonMachet2} for the decay of the neutral pion into two gauge
fields).

\subsection{A remark on the $\xi$ field.}

In the abelian case, one could think that nature is poorly described since
$\xi$, being absorbed by the massive gauge boson, acquires the same mass,
and that the old mass problem which gave rise to `technicolour'
\cite{SusskindWeinberg} theories, springs again.
This is however not the case in the $SU(2)_L\times U(1)$ case, as shown in
\cite{BellonMachet2}, because the fields to be absorbed by the
massive gauge bosons are linear combinations of the
physical pseudoscalar mesons, controlled by the mixing matrix
\cite{KobayashiMaskawa}, and all contain one top quark at least.
So, in this framework, another scale of
energy would only be needed if the `topped' mesons were found in a range
totally different from that of the $Z$ and $W$ mass.

\subsection{More on the Nambu-Jona-Lasinio approximation and beyond.}

We have shown the relevance of a yet unexplored domain of the
Nambu-Jona-Lasinio mechanism corresponding to infinite bare fermion mass and
infinite bare 4-fermions couplings. Our goal is clearly less
ambitious than the original one in \cite{NambuJonaLasinio}, which aimed at
showing {\em ab initio} that 4-fermions couplings yielded the formation of
bound states; we have instead taken advantage of the algebraic structure
of the model (which reduces here to the condition (\ref{eq:assoc}), but will
take its full significance and peculiarity in the case of the Standard Model
\cite{BellonMachet2}) to deliberately choose a set of composite scalar fields
transforming in the right way, and have shown that, in this approximation,
the fermions making the bound states could not appear as `particles'  as
soon as the symmetry was broken by $\la H\ra = v$, or, equivalently, by
$\la\overline \Psi \Psi\ra = N\mu^3$. But this phenomenon of fermionic
condensation was also taken as granted, and we did not seek to relate it
either with the presence of 4-fermions couplings.

It is a natural concern to investigate next orders in the $1/N$ expansion;
this is currently under scrutiny
\cite{BellonMachet3}.

\section{Conclusion.}

This work, together with \cite{BellonMachet1,Machet1}, and \cite{BellonMachet2}
to come, opens the possibility that, at least in the
Nambu-Jona-Lasinio approximation, the leptonic and hadronic sectors of
spontaneously broken gauge theories can be made anomaly-free independently,
and disconnected. It also suggests that fermions may not appear as
asymptotic states for other reasons than peculiar infrared properties, and
that the scalar boson, the vacuum expectation value of which is responsible
for the breaking of the symmetry, might not be detectable.

We hope that those results are sufficiently new to trigger more detailed
investigations.

\vskip 1cm
\nopagebreak
{\em\underline {Acknowledgements}: I benefited from discussions with my
colleagues in LPTHE, and special thanks are due to M. Bellon, M. Talon,
and C. M.  Viallet  for helping me to improve the redaction of this work.}
\newpage\null
\listoffigures
\bigskip
\begin{em}
Fig.1: Cancellation of the ghost residue by the Goldstone in the Landau
gauge;\\
Fig.2: Resumming the $\sigma_\mu$ and $\xi$ propagators;\\
Fig.3: Other diagrams for ghost cancellation in a general $\alpha$-gauge;\\
Fig.4: The effective 4-fermion coupling $\zeta(q^2)$;\\
Fig.5: Resumming the fermion propagator;\\
Fig.6: Resumming the scalar propagator;\\
Fig.7: Triangular diagrams involved in the anomalous Ward Identity;\\
Fig.8: Possible contributions to counterterms that can be dropped;\\
Fig.9: Diagram that could cause double-counting problems.\\
Fig.10: Cancelling the derivative coupling of $\xi$ to the fermions.

\newpage\null
%
%

\end{em}

\end{document}